\documentclass[10pt]{article}
\usepackage{graphicx,amsfonts}
\def\be{\begin{eqnarray}}
\def\ee{\end{eqnarray}}
\def\ds{\displaystyle}

\newcommand{\mapright}[1]{%
   \smash{\mathop{%
   \hbox to 1cm{\rightarrowfill}}\limits^{#1}}}
\newcommand{\mapleft}[1]{%
   \smash{\mathop{%
   \hbox to 1cm{\leftarrowfill}}\limits^{#1}}}
\newcommand{\maplleft}[2]{%
   \smash{\mathop{%
   \hbox to 1cm{\leftarrowfill}}\limits_{#1}^{#2}}}

\begin{document}

\title{Exact shock solution of a coupled system of delay differential equations: a car-following model }
\author{Y Tutiya and M Kanai}

\date{}
\maketitle
\begin{center}
{Graduate School of Mathematical Sciences\footnote{TEL:+81-3-5465-7001,FAX:+81-3-5465-7012\\
EMAIL:tutiya@poisson.ms.u-tokyo.ac.jp\, \,kanai@ms.u-tokyo.ac.jp}, The University of
Tokyo, Komaba 3-8-1, Meguro-ku, Tokyo 153-8914, Japan}\\

\end{center}

\begin{abstract}
In this paper, we present exact shock solutions
 of a coupled system of delay differential equations,
 which was introduced as a traffic-flow model
  called {\it the car-following model}.
We use the Hirota method, originally developed
 in order to solve soliton equations.
The relevant delay differential equations have been known
 to allow exact solutions expressed by elliptic functions
 with a periodic boundary conditions.
In the present work, however, shock solutions are obtained with open boundary, representing the stationary propagation of a traffic jam.
\end{abstract}


\section{Introduction}

Studies of traffic flow including pedestrian flow
 connect with a wide range of social problems such as
 efficient transport, evacuation in case of emergency and city planning.
One can regard traffic flows as compressible fluids
 from a macroscopic viewpoint,
 or as a many-body systems of driven particles
 from a microscopic viewpoint.
Accordingly, several different types of models have appeared; some are based
 on hydrodynamic equations,
 some are coupled differential equations,
 and others are cellular automata.

\begin{figure}[hbt]
\begin{center}
\includegraphics[scale=0.45]{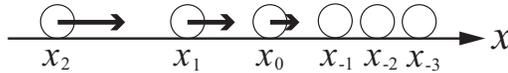}
\caption{An illustration of one-lane traffic.
It is prohibited for particles to overtake and collide.}
\label{lane}
\end{center}
\end{figure}

In particular, highway-like traffic is modeled
 as a one-dimensional system where a number of particles
 move in a definite direction interacting with each other
 asymmetrically.
(See Fig.\ref{lane}.)
The one-dimensional flow presents contrasting density patterns,
 which change quite irregularly
 as the density of particles increases,
 and which finally take the form of a stable traffic jam
 propagating backwards with constant speed.
In earlier work on traffic flow,
 {\it car-following models}
 were introduced in the form of a coupled system of DDEs:
\begin{equation}
\dot{x}_n(t)=F(\Delta x_n(t-\tau))\qquad(n=1,2,\ldots),
\label{cf}
\end{equation}
 where $x_n(t)\ (n=1,2,\ldots)$ denotes
 the position of the $n$th car
 at time $t$, $\Delta x_n(t)=x_{n-1}(t)-x_{n}(t)$
 is the distance
 to the next car in front (i.e. to the $(n-1)$th car),
 and $F$ is a given function.
The car-following model defined by (\ref{cf}) describes a situation where  each car determines its velocity $\dot{x}_n(t)$
 in terms of the distance the distance that separates it from its predecessor with a delay $\tau$, i.e., in terms of $\Delta x_n(t-\tau)$.
The delay $\tau$ is small but is never negligible,
 as the driver needs time to respond to the changing traffic situation,
 and hence it is one of the essential elements
 in traffic modeling.
The function $F$, which can be determined from real traffic data, provides the optimal velocity for each distance
 to which the drivers adjust their speed.
We call it {\it the optimal velocity (OV) function}
 hereafter.

In this paper, we present exact solutions for
 the car-following models with OV functions of exponential type (the Newell model) and $\tanh$ type (the so-called $\tanh$ model).
Using the Hirota method, which was originally developed for solving soliton equations, 
we obtain shock solutions representing a traffic jam that propagates backwards.

The Newell and the $tanh$ model have been studied extensively in the past and some exact solutions, typically involving elliptic functions, have been obtained.
However, all these solutions were obtained by assuming a specific traveling wave Ansatz, which reduces the original partial DDE to an ordinary DDE.
But, as the lattice intervals of $t$ and $n$ are different in the original partial DDE, the reduced ordinary equations inevitably include two incommensurate lattice intervals, which makes the problem extraordinarily difficult.
To solve this situation, the previous authors always required a strict relation between the velocity of the traveling wave and the delay (it must be equal to $2\tau$). This requirement then reduces the problem to a nonlinear ordinary difference equation of second order (with a first order differential term), which can be associated with addition theorems for the elliptic functions. 
It seems that up to now this was the only method available to tackle such nonlinear delay equations.
However, in the present paper, we shall demonstrate that Hirota's method is very well suited to treat such equations. 
Exactly how this method can be applied for solving  (\ref{cf}) will be shown in the subsequent sections.

The main results of this paper are reported in Section II and III.
In Section II, the Newell model will be introduced and its bilinear form and a shock solution will be presented. 
Section III is dedicated to the $\tanh$ model,
where as in Section II, a shock solution is obtained from the bilinear formalism.
Section IV is devoted to conclusions and remarks.

\section{The Newell model}
In the epoch-making work \cite{Newell},
 G F Newell proposed a car-following model
 described by the DDE
\begin{equation}
\dot{x}_n(t)=V\left[1-\exp\left(-\frac{\gamma}{V}(\Delta x_n(t-\tau)-L)\right)\right].
\label{NE}
\end{equation}
We call it {\it the Newell model}
 and (\ref{NE}) {\it the Newell equation} hereafter.
In \cite{Newell}, particular solutions for (\ref{NE}) in the case $\tau=0$ were presented.
Also, in the subsequent work by G B Whitham \cite{Whitham},
  exact solutions representing solitary waves
 and periodic waves were found for the case $\tau\ne0$.

\begin{figure}[hbt]
\begin{center}
\includegraphics[scale=0.7]{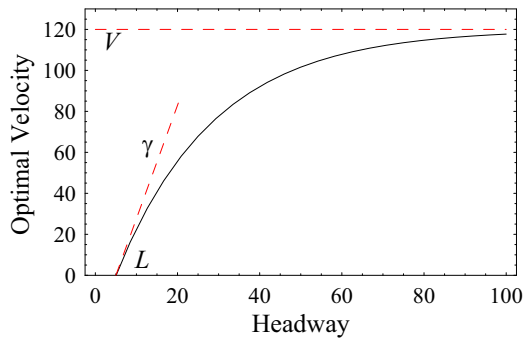}
\caption{The optimal velocity function for the Newell model.
The values of the parameters are as follows:
 $V=120$, $\gamma=6$ and $L=5$,
 where $V$ indicates the maximum allowed velocity,
 $\gamma$ is the derivative of $F$ at $\Delta x_n=L$,
 and $L$ is the minimum distance
 between cars.}
\label{NOV}
\end{center}
\end{figure}

The parameters in (\ref{NE}) should be interpreted as follows:
 $V$ is the maximum allowed velocity of a car, 
 $\gamma$ (the slope of the OV function at $\Delta x_n=L$) corresponds
 to the sensitivity of the driver to changes in the traffic situation,
 and $L$ is the minimum headway.
Fig.\ref{NOV}
 shows the graph of the OV function.
The parameters can be determined from empirical data,
 and in \cite{Newell} the Newell model,
 for certain estimated parameter values,
 was shown to give a reasonable fit of the empirical data.

Firstly, eliminating uniform flows
 for the sake of convenience, we change the dependent variable
 from $x_n$ to $y_n$ as
\begin{equation}
x_n(t)=V_0t-L_0n+y_n(t),
\label{xy}
\end{equation}
 where the velocity $V_0$ and headway $L_0$ satisfy the relation
\begin{equation}
V_0=V\left[1-\exp\left(-\frac\gamma V(L_0-L)\right)\right]
\label{VV}
\end{equation}
that is required of a uniform solution.
Then, substituting (\ref{xy}) in (\ref{NE})
 and using (\ref{VV}), the Newell model becomes:
\begin{equation}
\dot{y}_n(t)=(V-V_0)\left[1-\exp\left(-\frac\gamma V\Delta y_n(t-\tau)\right)\right].
\label{yNE}
\end{equation}
Considering differences of (\ref{yNE}), setting 
$s_n(t)=(\gamma/ V)\{y_{n-1}(t)-y_n(t)\}$,
one obtains the following equation,
\begin{equation}
\frac1{\alpha_0}\dot{s}_n(t)=-\exp\left(-s_{n-1}(t-\tau)\right)+\exp\left(-s_{n}(t-\tau)\right),
\label{sss}
\end{equation}
where $\alpha_0=\gamma(1-V_0/V)=\gamma\exp(-(\gamma/V)(L_0-L))$.

For instance, (\ref{sss}) is known to have the following elliptic solution \cite{Whitham}:
\be
&&s_n(t)=\log\frac{2\alpha_0{\rm sn}(\Omega\tau){\rm cn}(\Omega\tau){\rm dn}(\Omega\tau)}{\Omega\left(1-k^2{\rm sn}^2(\Omega\tau){\rm sn}^2(\phi+(\Omega\tau))\right)\left(1-k^2{\rm sn}^2(\Omega\tau){\rm sn}^2\phi\right)},
\label{S}
\ee
 where $\phi =\Omega(t-2\tau n),\ {\rm sn}\phi={\rm sn}(\phi;k),\ {\rm cn}\phi={\rm cn}(\phi;k)$,
 and ${\rm dn}\phi={\rm dn}(\phi;k)$ are Jacobi's elliptic functions
 with modulus $k$.
$\Omega$ is determined by a certain transcendental equation.

\subsection{A shock solution with parameteric freedom}
Now we apply the Hirota method \cite{Hirota1,Hirota2} to the Newell model.
In order to rewrite (\ref{sss}) as a DDE in algebraic form,
 we first set
$\psi=\exp(-s_{n}(t))$.
Then, (\ref{sss}) transforms 
 into an algebraic relation between $\psi$ and $\dot{\psi}$:
\begin{equation}
\frac1\alpha_0\frac{\dot{\psi}_n(t+\tau)}{\psi_n(t+\tau)}=\psi_{n-1}(t)-\psi_n(t).\label{witham2}
\end{equation}
Following the examples of soliton equations \cite{Hirota1,Hirota2}, 
we set
$\psi=g/f$
, where $f$ and $g$ are usually considered to be entire functions.
Then, (\ref{witham2}) can be written in the following form.
\begin{equation}
\frac{1}{\alpha_0}\frac{\dot{g}^+f^+-g^+\dot{f}^+}{f^+g^+}=\frac{\underline{g}f-g\underline{f}}{\underline{f}f},\label{withm3}
\end{equation}
where for simplicity we use the notations, $\underline{f}=f(n-1,t)$ and $f^\pm=f(n,t\pm\tau)$ to denote $n$- and $t$-shifts, respectively.
As a result, (\ref{withm3}) can be decoupled
 into the following system:
\begin{equation}
\left\{
\begin{array}{l}
\dot{g}^+f^+-g^+\dot{f}^+=\lambda(\underline{g}f-g\underline{f})\\
\ds\underline{f}f=\frac{\alpha_0}{\lambda} f^+g^+,
\end{array}
\right.\label{witham4}
\end{equation}
 where $\lambda$ is introduced as a coupling constant.
This is the so-called bilinear form of (\ref{NE}).

Next, we assume that $f=1+\exp(an+2bt)$ and $g=u+v\exp(an+2bt)$,
 where $a,\,b,\,u$, and $v$ are constants
 and the factor $2$ in front of $b$ is for later convenience.
By substituting $f$ and $g$ into (\ref{witham4}),
 the coefficients of both exponentials
 give rise to the constraints on the constants.
In this case, the number of constraints is less than
 that of the constants and $f$ and $g$
 can be determined as follows:
\begin{equation}
\left\{
\begin{array}{l}
f=1+\exp(2b(t-\tau n))\\
\displaystyle 
g=\frac{b}{\alpha_0(1-e^{-2b\tau})}\{1+\exp(2b[t-\tau(n+1)])\},
\end{array}
\right.\label{witham6}
\end{equation}
 where $b$ is a free parameter.
We thus obtain an exact solution of (\ref{NE}):
\begin{equation}
s_n(t)=\log\frac{\alpha_0\sinh (b\tau)}{b}\frac{\cosh\left(b(t-\tau n)\right)}{\cosh(b[t-\tau(n+1)])}.
\label{witham7}
\end{equation}
This is a shock wave with velocity $U=1/\tau$
 which presents a traffic jam propagating backwards;
 and accordingly open boundary conditions are assumed
(or equivalently, the period of a wave type solution is assumed to be infinite).
It is remarkable that the shock velocity $U$
 (a macroscopic value) should be determined by
 the driver's response time $\tau$ (a microscopic value).
At the same time, this condition on the velocity means that this solution does not reduce to any previously known solution for which one always had $U=1/2\tau$.

The previous solution (\ref{S}) is recovered if one takes
\begin{equation}
f=\vartheta_0^+(\phi)\vartheta_0(\phi),\quad g=\vartheta_0^{++}(\phi)\vartheta_0^-(\phi),\label{fg1}
\end{equation}
where $\vartheta_0(\phi)=\vartheta_0(\phi;k)$ denotes
 the elliptic theta function with modulus $k$.

%

%
\section{Car-following model with the tanh OV function}
In \cite{IIN,HNS}, two groups of authors independently found
 that yet another car-following model allows for exact solutions,
 also described by elliptic functions.
This model is represented by the DDE
\begin{equation}
\dot{x}_n(t)=\xi+\eta\tanh\left(\frac{\Delta x_n(t-\tau)-\rho}{2A}\right),\label{TCF}
\end{equation}
where $\xi,\ \eta,\ \rho$ and $A$ are parameters.
We call the car-following model with the OV function (\ref{TCF})
 {\it the tanh car-following model} (or simply, tanh model),
 hereafter.

The tanh OV function (the RHS of (\ref{TCF}))
 was extensively investigated
 in the context of the OV model \cite{Bando1,Bando2}.
In particular, numerical simulation showed that this OV function reproduces the onset and growth of traffic jams faithfully.

For convenience and simplicity, we shift the delay
 from the RHS to the LHS in (\ref{TCF}),
 and introduce the distance variables
$h_n(t)=(\Delta x_n(t)-\rho)/2A$
 instead of the positions $x_n$.
Then, the tanh car-following model is represented
 by the equation:
\begin{equation}
\dot{h}_n(t+\tau)=\frac{\eta}{2A}\left[\tanh h_{n-1}(t)-\tanh h_n(t)\right].\label{h}
\end{equation}
As for the equation (\ref{h}), some elliptic solutions are known \cite{IIN,HNS}.
Typical examples can be written in the following forms:
\be
h_n(t)=a\ {\rm sn}(\phi)+b,\quad \frac{b}{{\rm sn}(\phi)+a}+c,\quad \frac{b}{{\rm sn}^2(\phi)+a}+c,\label{sns}
\ee
where $\phi=\Omega(t-2\tau n)$ and $\Omega$ can be regarded as a free parameter.
$a,b,c$ are constants determined by $\Omega, A, \eta$ and $\tau$.
\subsection{A shock solution for the tanh model}
In terms of $\psi=\tanh h_n$, (\ref{h}) takes the form,
 into an algebraic form
\begin{equation}
\dot{\psi}^+=\frac{\eta}{2A}\left\{1-(\psi^+)^2\right\}(\underline{\psi}-\psi)\label{tanh2}.
\end{equation}
It should be noted that, if $\psi$ is a solution of (\ref{tanh2}), $-\psi$ is also a solution.
As was the case of the Newell model,
 we set $\psi=g/f$,
 and thereby successfully separate (\ref{tanh2})
 into a system of bilinear equations:
\begin{equation}
\left\{
\begin{array}{l}
\dot{g}^+f^+-g^+\dot{f}^+=\lambda(\underline{g}f-g\underline{f})\\
\displaystyle\underline{f}f=\frac{\eta}{2A\lambda}[(f^+)^2-(g^+)^2],
\end{array}
\right.
\label{tanh4}
\end{equation}
 where $\lambda$ is a coupling constant.
This is the bilinear form of (\ref{h}).

Substituting $f=1+\exp(2bt-an)$ and $g=u+v\exp(2bt-an)$
 into (\ref{tanh4}), we find
\begin{equation}
\left\{
\begin{array}{l}
f=1+{\rm exp}(2bt-an)\\
\displaystyle g=\left(1-\frac{2bA}{\eta(1-e^{-2b\tau})}\right)\left[1+e^{-2b\tau}\exp(2bt-an)\right],
\end{array}
\right.\label{bl}
\end{equation}
where $a$ is given by
\begin{equation}
e^a=\frac{bA/\eta+1-e^{2b\tau}}{bA/\eta-1+e^{-2b\tau}}.
\label{d}
\end{equation}
{From} (\ref{bl}), we finally obtain the following exact solution of (\ref{TCF}):
\begin{eqnarray}
\Delta x_{n}(t)=\rho\nonumber+A\log\left(\frac{2\eta\sinh (b\tau)}{bA}\frac{\displaystyle\cosh\left(bt-\frac{a}{2}n\right)}{\displaystyle\cosh\left(b(t-\tau)-\frac{a}{2}n\right)}-1\right).
\label{shock2}
\end{eqnarray}
This is another shock wave with velocity $U=2b/a$
 which represents a traffic jam propagating backwards.
It should be noticed that the velocity of (\ref{shock2}) is free, whereas the velocity of the solutions previously reported are all fixed at $1/2\tau$ \cite{IIN,HNS}.

The previous solutions (\ref{sns}) are recovered if one assume $f$ and $g$ are the following forms:
\be
f^+(\phi)=J\vartheta_1(\phi)+K\vartheta_0(\phi),\quad
g^+(\phi)=L\vartheta_1(\phi)+M\vartheta_0(\phi),
\label{a1}
\ee
(which correspond to solutions of the type 
$s_n(t)=a\ {\rm sn}(\phi)+b$ and $b/({\rm sn}(\phi)+a)+c$,) or
\be
f^+(\phi)=J\vartheta_1^2(\phi)+K\vartheta_0^2(\phi),\quad
g^+(\phi)=L\vartheta_1^2(\phi)+M\vartheta_0^2(\phi),
\label{a2}
\ee
(which correspond to solutions of the type $b/({\rm sn}^2(\phi)+a)+c$).
For the solution (\ref{a1}), the constants $J,K,L,M$ are determied by following quadratic relations:
\be
\left\{
\begin{array}{l}
\ds\mu=\frac{\Omega A\vartheta_0\vartheta'_1}{\eta\tau \vartheta_0(\Omega)\vartheta_1(\Omega)}\\[3mm]
\ds L^2=\left(1-\mu \frac{\vartheta_0^2(\Omega)}{\vartheta_0^2}\right)J^2+\mu \frac{\vartheta_1^2(\Omega)}{\vartheta_0^2}K^2\\[3mm]
\ds LM=\left(1-\mu \frac{\vartheta_2(\Omega)\vartheta_3(\Omega)}{\vartheta_2\vartheta_3}\right)JK\\[3mm]
\ds M^2=\mu \frac{\vartheta_1^2(\Omega)}{\vartheta_0^2}J^2+\left(1-\mu \frac{\vartheta_0^2(\Omega)}{\vartheta_0^2}\right)K^2
\end{array}\right..
\ee
For the solution (\ref{a2}), the constants are determied by following quartic relations:
\be
\left\{
\begin{array}{l}
\ds\mu=\frac{\Omega A\vartheta_0\vartheta'_1\vartheta_2\vartheta_3}{\eta\tau \vartheta_0(\Omega)\vartheta_1(\Omega)\vartheta_2(\Omega)\vartheta_3(\Omega)}
\\[3mm]
\ds
L^2=\left(1-\mu\frac{\vartheta_0^4(\Omega)}{\vartheta_0^4}\right)J^2
-\mu\frac{\vartheta_1^4(\Omega)}{\vartheta_0^4}K^2
-2\mu\frac{k}{{k'}^2}\frac{\vartheta_0^2(\Omega)\vartheta_1^2(\Omega)}{\vartheta_2^2\vartheta_3^2}JK
\\[3mm]
\ds
LM=
\mu\frac{\vartheta_0^2(\Omega)\vartheta_1^2(\Omega)}{\vartheta_0^4}J^2
+\mu\frac{\vartheta_0^2(\Omega)\vartheta_1^2(\Omega)}{\vartheta_0^4}K^2
+\left(1+\mu\frac{1+k^2}{{k'}^2}\frac{\vartheta_0^2(\Omega)\vartheta_1^2(\Omega)}{\vartheta_2^2\vartheta_3^2}\right)JK
\\[3mm]
\ds
M^2=-\mu\frac{\vartheta_1^4(\Omega)}{\vartheta_0^4}J^2+
\left(1-\mu\frac{\vartheta_0^4(\Omega)}{\vartheta_0^4}\right)K^2
-2\mu\frac{k}{{k'}^2}\frac{\vartheta_0^2(\Omega)\vartheta_1^2(\Omega)}{\vartheta_2^2\vartheta_3^2}JK
\end{array}\right..
\ee
In above relations and , $\vartheta_j(\phi)=\vartheta_j(\phi;k)$ denotes
 the elliptic theta function with modulus $k$ and $\vartheta_j$ denotes $\vartheta_j(0)$.
\section{Summary and Conclusion}
In this paper, we presented the Hirota bilinear forms of two typical car-following models, and obtained exact shock solutions.
These are the first solutions that do not assume the velocity of the traveling  wave to be equal to $1/2\tau$.

The Hirota method was originally introduced to find special
solutions such as soliton solutions for nonlinear partial differential or difference equations.
We believe that the method is effective for solving nonlinear equations in general and delay-differential equations in particular.

\end{document}